\documentclass[prb,amsmath,superscriptaddress,showpacs,showkeys,prapplied,twocolumn]{revtex4-2}
\usepackage{amssymb,amsmath}   
\usepackage[dvips]{graphicx}   
\usepackage{verbatim}   
\usepackage{color}      
\usepackage{hyperref}   
\usepackage{gensymb}
\usepackage{epstopdf}
\usepackage{natbib}
\usepackage{braket}
\usepackage{enumerate}

\begin{document}

\title{Qubit dephasing by spectrally diffusing quantum two-level systems}
	
\author{Shlomi Matityahu}\email{shlomimatit@gmail.com}
\affiliation{Institut f\"ur Theorie der Kondensierten Materie, Karlsruhe Institute of Technology, 76131 Karlsruhe, Germany}
\author{Alexander Shnirman}
\affiliation{Institut f\"ur Theorie der Kondensierten Materie, Karlsruhe Institute of Technology, 76131 Karlsruhe, Germany}
\affiliation{Institut f\"ur Quantenmaterialien und Technologien, Karlsruhe Institute of Technology, 76021 Karlsruhe, Germany}
\author{Moshe Schechter}
\affiliation{Department of Physics, Ben-Gurion University of the Negev, Beer Sheva 84105, Israel}

\date{\today}
\begin{abstract}
	We investigate the pure dephasing of a Josephson qubit due to the spectral diffusion of two-level systems that are close to resonance with the qubit.
	We identify the parameter regime in which this pure dephasing rate can be of the order of the energy relaxation rate and, thus, the relation $T_2 = 2 T_1$ is violated for the qubit. This regime is reached if the dynamics of the thermal TLSs responsible for the spectral diffusion is sufficiently slower than the energy relaxation of the qubit. By adding periodic bias modulating the qubit frequency or TLS excitation energies we show that this pure dephasing mechanism can be mitigated, allowing enhancement of superconducting qubits coherence time.
Mitigating pure dephasing, even if it is subdominant, is of special significance in view of recent suggestions for converting the dominant relaxation process ($T_1$) into erasure errors, leaving pure dephasing as the bottleneck for efficient quantum computation.
\end{abstract}

\date{\today}

\maketitle \pagenumbering{roman} \pagenumbering{arabic}

\section{Introduction}

Superconducting Josephson qubits have made an impressive impact on the field of quantum  artificial systems in general and on quantum computing in particular~\cite{QubitsReview2020}. Improving the coherence and stability of these qubits is necessary to make further progress. This would allow increasing the computational depth, reducing the unavoidable errors and making more reliable quantum simulations. In this paper we investigate one of the main sources of dissipation in Josepshon quantum circuits - atomic defects with a two-level structure, that act as parasitic "hidden" qubits and couple to the logical qubits via the circuit charge or flux degrees of freedom. Such environmental exciatations are commonly referred to as two-level systems (TLSs) in the literature.

The TLSs with energy splitting of the order of that of the qubit are considered to be one of the major sources of qubit energy relaxation ($T_1$-process). Namely, the qubit can exchange an excitation with such a TLS and the TLS, in turn, dissipates its energy to a thermal bath of phonons. This mechanism has been investigated in great detail~\cite{Martinis2005,KlimovPRL18,Lisenfeld2019,Abdurakhimov2022}. For clarity we will call such TLSs with energy splitting close to that of the qubit "quantum TLSs".

\begin{figure}
\includegraphics[width=0.5\textwidth,height=5cm]{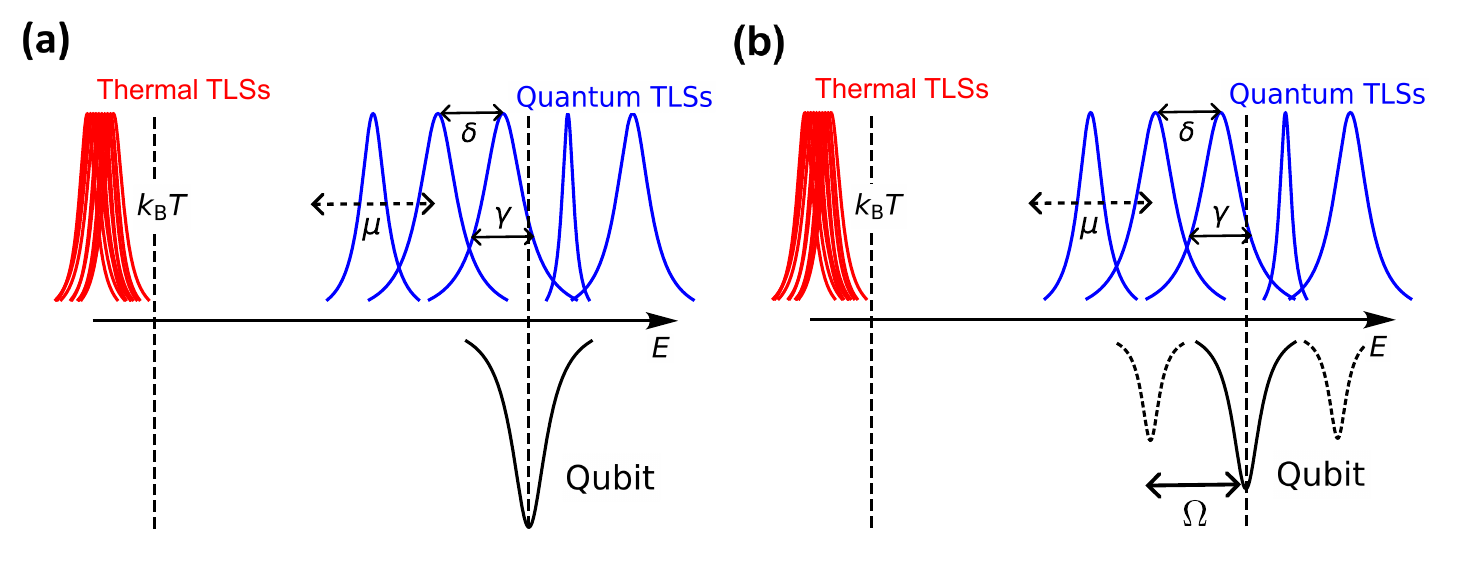}
\caption{Spectral schematics of the qubit, the quantum TLSs performing spectral diffusion and the thermal TLSs responsible for it. Each quantum TLS is
characterized by an absorption peak (blue) of width $\gamma$. The quantum TLSs perform spectral diffusion of span $\mu$, which is caused by coupling to the thermal TLSs (red). The typical energy distance between adjacent quantum TLSs is $\delta$. a) without periodic modulation; b) with periodic modulation with frequency $\Omega$. In this case the absorption peak of the qubit is split into multiple peaks at spectral distance $\Omega$.}
\label{fig:TLS}
\end{figure}

In addition, it is known that frequently randomly looking ''$T_1$ fingerprint'', i.e. the dependence of $T_1$ on the qubit frequency, fluctuates in time~\cite{KlimovPRL18,SchloerPRL19,BurnettNPJQI19,Carroll2022}. This can be explained by the spectral diffusion of the TLSs, i.e. the random diffusive motion of the TLS energy splitting~\cite{KlauderAndersson62}. This diffusion is explained by coupling of the quantum TLSs to an ensemble of thermal TLSs, i.e. TLSs with energy splittings smaller or of the order of the temperature. These thermal TLSs switch randomly between their two states, thus shifting the energy of the quantum TLSs (Fig.~\ref{fig:TLS}a).

Coupling of the qubit to the quantum TLSs causes not only energy relaxation but also frequency shift, known as the Lamb shift. The two are intimately related via the Kramers-Kronig relations. Spectral diffusion gives rise to temporal fluctuations of the Lamb shift and, thus, to the pure dephasing of the qubit. In this paper we investigate this process in detail and identify regimes in which the pure dephasing time is of the order of the relaxation time $T_1$. Following our previous paper~\cite{MatityahuShnirmanSchechter21}, we also investigate the effect of qubit frequency modulation on its pure dephasing time.

We index the quantum TLSs by an integer $n$. Each quantum TLS is characterized by several parameters: $\epsilon_n$ - the energy splitting of the TLS,
$g_n$ - the coupling strength of the TLS to the qubit, $\gamma_n$ - the relaxation rate of the TLS due to phonons. The spectral diffusion adds yet another parameter $\mu_{\rm{av},n}$, which gives the characteristic span of the spectral diffusion in the long-time limit. In other words, this is the width of the energy interval "visited" by $\epsilon_n(t)$ over long times. In addition, the ensemble of the quantum TLSs is characterized by the typical level spacing $\delta$, namely the energy distance, $\epsilon_n - \epsilon_{n+1}$, between spectrally neighboring TLSs (see Fig.~\ref{fig:TLS}a). Taking into account the wide distribution of the coupling strengths $g_n$, it might be more prudent to consider separately the subsets of quantum TLSs with a specific coupling strength and introduce the $g$-dependent level spacing $\delta(g)$ as was done in Ref.~\cite{Martinis2005}.

Historically, the first TLSs observed spectroscopically by Simmonds et al. in 2004~\cite{Simmonds2004} were in the regime of very strong coupling $g  \gg \gamma$ and spectrally rare $\delta \gg g$ (we drop the index $n$ for brevity). The coupling strength $g$ was observed as the avoided level crossing in qubit spectroscopy. The spectral diffusion did not play any significant role, and one could estimate $\mu_{\rm{av}} \ll \delta$ and even
$\mu_{\rm{av}} \ll g$.

Similarly, spectrally rare ensembles of TLSs strongly interacting with the qubit were observed in later works~\cite{Martinis2005,Grabovskij2012}. In more recent studies~\cite{Barends2013,KlimovPRL18,Lisenfeld2019}, denser ensembles of quantum TLSs with weaker coupling to the qubit ($g<\gamma$) were observed.
Such TLSs cause peaks in the relaxation rate of the qubit as a function of its energy splitting, denoted henceforth $T^{-1}_{1,\rm{q}}$. The randomly looking dependence of the qubit relaxation rate on the qubit frequency suggested that $\delta \sim \gamma$, i.e., $T^{-1}_{1,\rm{q}}$ was given by multiple partly overlapping Lorentzians. Moreover, following the time-dependence of the positions of the strongest peaks~\cite{KlimovPRL18}, one could conclude that $\mu_{\rm{av}} \gg \gamma$.

Here we propose a more differentiated approach. The experiments~\cite{Barends2013,KlimovPRL18,Lisenfeld2019} show that the strongest peaks are still relatively rare, such that the level spacing of these peaks satisfies $\delta > \mu_{\rm{av}}  >  \gamma$.  These are relatively sparse TLSs that are strongly coupled to the qubit and at the same time are close to resonance with it. However, these peaks are not responsible for the background level of
$T^{-1}_{1,\rm{q}}$. We follow Ref.~\cite{Martinis2005} and assume the existence of a much denser subset of quantum TLSs with an even weaker coupling to the qubit, a result of the abundance of TLSs with small tunnelling amplitudes. These TLSs are so spectrally dense, such that the span of their spectral diffusion is larger than their level spacing and their inverse life time, i.e. $\mu_{\rm{av}}  >  \gamma,\delta$.  It is this subset which is mostly influenced by the spectral diffusion and can give the strongest contribution to the pure dephasing of the qubit. In the other regimes, e.g. for $\mu_{\rm{av}}  < \gamma$, dephasing is suppressed as discussed in Appendix~\ref{Sec:SmallDiffusionLimit}.

To be concrete we assume, following Ref.~\cite{Martinis2005}, that the TLS density of states $dN/d\varepsilon$ satisfies
\begin{align}\label{eq:Distributionofg}
\frac{d^2 N}{dg d\varepsilon} = \delta_{\rm typ}^{-1} \frac{\sqrt{1-g^2/g_{\rm max}^2}}{g}\quad{\rm for}\quad g<g_{\rm max}
\end{align}
and zero otherwise. Here $\delta_{\rm typ}$ is a constant depending on the materials and the geometric shape of the qubit. It will play the role of the typical level spacing as we will see below. We conjecture this distribution to be valid down to very small values of $g$. Then it is guaranteed that we have a large number of TLSs such that $\mu_{\rm{av}}  >  \gamma,\delta$ at high enough temperatures (note that $\mu_{\rm av}\propto T$, see Ref.~\cite{BlackHalperin77}). These are the TLSs which are of interest to us in this paper.

Analyzing the pure dephasing of the qubit due to the spectral diffusion, we will see that a major role is played by the typical relaxation time of the thermal TLSs, which we denote by $T_{1,\rm{T}}$. This is the time scale at which the spectral diffusion of a quantum TLS saturates at the span of order $\mu_{\rm{av}}$. Our main result is the pure dephasing law $D(t)$, which reads
\begin{widetext}
\begin{align}\label{eq:MainResultIntro}
	-2 \ln D(t)
	\sim\frac{g_{\rm max}^{4}}{\delta_{\rm typ}}\begin{cases}
		\frac{1}{\gamma}t^2, \quad t\ll\frac{\gamma}{\mu_{\rm{av}}}T_{1,\rm{T}}\\
		\frac{T_{1,\rm{T}}}{\mu_{\rm{av}}}t\ln\left(\frac{\mu_{\rm{av}}}{T_{1,\rm{T}}\gamma}t\right), \quad  T_{1,\rm{T}} \gg t\gg\frac{\gamma}{\mu_{\rm{av}}}T_{1,\rm{T}}\\
		\frac{T_{1,\rm{T}}}{\mu_{\rm{av}}}t\ln\left(\frac{\mu_{\rm{av}}}{\gamma}\right), \quad t\gg T_{1,\rm{T}}
	\end{cases}.
\end{align}
\end{widetext}
Here, $\gamma$ and $\mu_{\rm av}$ stand for typical values (see Appendix~\ref{Sec:EstimatingDt}).
This result is valid at least as long as $-2 \ln D(t) \lesssim 1$, thus it can be safely used to estimate the effective dephasing rate (time). In Sec.~\ref{sec: discussion} we conclude that at low enough temperatures, when $T_{1,\rm{T}} /T_{1,\rm{q}} \gg  \mu_{\rm{av}}/\sqrt{\gamma\delta_{\rm typ}}$, the effective pure dephasing time due to the spectral diffusion becomes comparable to $T_{1,\rm{q}}$. This can explain numerous experimental observations, in which the qubit dephasing time $T_{2,\rm{q}}$ is significantly shorter than its upper limit $2T_{1,\rm{q}}$ (see, e.g., Refs.~[\onlinecite{PaikPRL2011,SchloerPRL19,BurnettNPJQI19,KandalaNature2019}]). Below we reanalyse the dephasing law (\ref{eq:MainResultIntro}) in the presence of periodic modulations of the qubit frequency and conclude that in this regime the modulations can strongly suppress the pure dephasing and, ultimately, restore the relation $T_{2,\rm{q}} = 2T_{1,\rm{q}}$. 

\section{Theoretical framework}
\label{sec: theory}
\subsection{The model}
\label{sec: model}
We consider the same Hamiltonian as in our previous paper~\cite{MatityahuShnirmanSchechter21},
\begin{align}
	\label{eq:eq1}\mathcal{H}=&-\frac{1}{2}\left[E_0+A\cos(\Omega t)\right]\sigma_z-\frac{1}{2}\sum^{}_{n}\varepsilon_n(t)\tau^{(n)}_{z}\nonumber\\
	&+\frac{1}{2}\sum^{}_{n}g_n\sigma_x\tau^{(n)}_{x}-\frac{i}{2}\sum^{}_{n}\gamma_n(1-\tau^{(n)}_{z})\ ,
\end{align}
where $\sigma_\alpha$ and $\tau^{(n)}_{\alpha}$ ($\alpha=x, y, z$) are the Pauli matrices that describe the qubit and the $n$th quantum TLS, respectively. The qubit energy splitting is modulated harmonically around $E_0$ with amplitude $A$ and frequency $\Omega$. In our previous paper~\cite{MatityahuShnirmanSchechter21} we assumed that the energy splittings $\varepsilon_n$ of the TLSs are constant (but randomly distributed). Here we consider the effect of spectral diffusion due to interaction with thermal TLSs, giving rise to fluctuations in $\varepsilon_n$. The third term is the interaction between the qubit and the TLSs, characterized by the coupling constants $g_n$. Finally, the dissipation of the TLSs due to phonons can be described by the last non-Hermitian term with relaxation rates $\gamma_n$. The reason this description is sufficient and the full Linbladian is not needed is explained in Appendix~\ref{Sec:NHHamiltonian} where the microscopic derivation is provided.

In our previous paper~\cite{MatityahuShnirmanSchechter21} we have studied the effect of the periodic modulation with frequency $\Omega$ on the energy relaxation rate of the qubit ($T^{-1}_{1,\rm{q}}$). The main effect was the replication of the absorption peaks in $T^{-1}_{1,\rm{q}}(E_0)$, at energies shifted from $E_{0}$ by multiples of $\Omega$ (see Fig.~\ref{fig:TLS}b). We have assumed the condition $\Omega> \gamma_n, g_n$, meaning that the replications of each absorption peak extend beyond its width and demonstrated that the modulation conserves the average of the relaxation rate over $E_{0}$ but the fluctuations in the relaxation rate can be suppressed. In this work we investigate the effect of the spectral diffusion on the pure dephasing of the qubit, both in the absence and in the presence of the periodic modulation in the regime $\Omega> \gamma_n, g_n$.

\subsection{Qubit relaxation and decoherence}
\label{sec: relaxation and decoherence}
We consider the initial state $\ket{\psi^{}_{}(t=0)} = (\alpha\ket{0} + \beta\ket{1}) \ket{0,\ldots,0}$, in which the qubit is prepared in a superposition of the ground ($\ket{0}$) and excited ($\ket{1}$) states, whereas all the quantum TLSs are in their ground states. Within the rotating wave approximation only the component containing the excited state of the qubit ($\propto\beta$) evolves in time in a non-trivial manner, whereas the component $\propto\alpha$ stays intact. As in our previous paper~\cite{MatityahuShnirmanSchechter21}, we analyze the model Hamiltonian~(\ref{eq:eq1}) using the Weisskopf-Wigner approach~\cite{WeisskopfWigner30}. We treat the third term as a perturbation, use the rotating-wave approximation, and consider the single-excitation ansatz $\ket{\psi^{}_{\mathrm{I}}(t)}=\alpha\ket{0}\ket{0,\ldots,0}+\beta\left[a(t)\ket{1}\ket{0,\ldots,0}+\sum^{}_{n}b_n(t)\ket{0}\ket{0,\ldots,0,1_n,0,\ldots,0}\right]$ for the state in the interaction picture, with the initial conditions $a(0)=1$, $b_{n}(0)=0\ \forall n$. The reduced density matrix of the qubit reads
\begin{align}\label{eq:densitymatrix}
	\rho = \left(
	\begin{array}{cc}
		1- |\beta|^2 |a(t)|^2 & \alpha \beta^* a(t)\\
		\alpha^* \beta a(t)^* &  |\beta|^2\,|a(t)|^2
	\end{array}
	\right)\ .
\end{align}
This expression should be averaged over the realizations of the spectral diffusion $\varepsilon_n(t)$, which will be denoted by $\langle\ \rangle$ (see Sec.~\ref{sec: Spectral diffusion} below). Below we solve the Schr\"{o}dinger equation to find the amplitude of staying in the excited state of the qubit, $a(t)$, and then consider its average $\langle a(t) \rangle$, as well the average $\langle|a(t)|^2\rangle$. From Eq.~(\ref{eq:densitymatrix}) we see that $\langle |a(t)|^2\rangle$ describes the relaxation, which we discussed in Ref.~\cite{MatityahuShnirmanSchechter21}, whereas $M(t)\equiv|\langle a(t) \rangle|$ contains also the pure dephasing, which is the subject of this paper.

The Schr\"{o}dinger (Weisskopf-Wigner) equations read~\cite{MatityahuShnirmanSchechter21}
\begin{align}
	\label{eq:eq2}
	&\dot{a}(t)=-\frac{i}{2}\sum^{}_{n}g_{n}e^{i\left[E_{0}t+\phi(t)-\int^{t}_{0}\varepsilon_n(t')dt'+i\gamma_{n}t\right]}b_n(t),\nonumber\\
	&\dot{b}^{}_{n}(t)=-\frac{i}{2}g_{n}e^{-i\left[E_{0}t+\phi(t)-\int^{t}_{0}\varepsilon_n(t')dt'+i\gamma_{n}t \right]}a(t),
\end{align}
where $\phi(t)=\int^{t}_{0}A\cos(\Omega t')dt'=\left(A/\Omega\right)\sin(\Omega t)$. We  integrate the second equation, employ the well-known expansion
$e^{i\phi(t)} = \sum_m J_m(A/\Omega) e^{im\Omega t}$ ($J_m(x)$ being the Bessel functions of the first kind), and substitute the result into the first equation. We obtain the following integro-differential equation
\begin{widetext}
\begin{align}
	\label{eq:eq3}&\dot{a}(t)=-\frac{1}{4}\sum^{}_{n}\sum^{}_{m^{}_{1},m^{}_{2}}g^{2}_{n}J_{m^{}_{1}}J_{m^{}_{2}}e^{i(m^{}_{1}-m^{}_{2})\Omega t}\int^{t}_{0}e^{i\left[\left(E_{0}+m_{2}\Omega\right)\tau-\int^{t}_{t-\tau}\varepsilon_{n}(t')dt'\right]}e^{-\gamma_{n}\tau}a(t-\tau)d\tau.
\end{align}
\end{widetext}
Here and in what follows we drop the argument $A/\Omega$ of the Bessel function.

In the weak coupling limit $g_{n}\ll\gamma_{n}$, we can employ the Markov approximation and replace $a(t-\tau)$ by $a(t)$ in the integrand. We can also extend the limit of the integration to $\infty$. The condition $\Omega\gg\gamma_n, g_n$ also implies that $e^{i(m_1-m_2)\Omega t}$ oscillates rapidly and averages to zero unless $m_1 = m_2$. If, in addition, $\varepsilon_{n}(t)$ varies slowly on the time scale of order $1/\gamma_{n}$, we can approximate $\int^{t}_{t-\tau}\varepsilon_{n}(t')dt'\approx\varepsilon_{n}(t)\tau$. This is justified if
\begin{align}\label{eq:T1Tcondition}
\int^{t}_{t-\tau}\left[\varepsilon_{n}(t')-\varepsilon_{n}(t)\right]dt'\ll 1
\end{align}
for $\tau \le 1/\gamma$,
i.e.\ if the maximum deviation in $\varepsilon_{n}(t)$ during the time interval of order $1/\gamma_n$ is smaller than $\gamma_n$. This condition [Eq.~(\ref{eq:T1Tcondition})] will be examined for self-consistency in Sec.~\ref{sec: discussion}. For the time being, we assume this condition holds and obtain the time-local differential equation $\dot{a}(t)=-C(t)a(t)$, where $C(t)=\sum_{n}C_n(t)$ and
\begin{align}
	\label{eq:eq4}
	C_n(t) = &\frac{1}{4}\sum^{\infty}_{m=-\infty}\frac{g^{2}_{n}J^{2}_{m}}{\gamma_n-i\left(E_0+m\Omega-\varepsilon_n(t)\right)}.
\end{align}
With the initial condition $a(0)=1$, the solution reads $a(t) = \prod_n a_n(t)$, where
\begin{align}
	\label{eq:eq5}
	a_n(t)=e^{-\int^{t}_{0}C_n(t')dt'}.
\end{align}
Since the spectral diffusion fluctuations of $\varepsilon_n(t)$ are independent, the averaging of $a(t)$ as well as of $|a(t)|^{2}$ can be performed separately for each $n$, i.e.
\begin{align}
	\label{eq:Dephasing}
	M(t)\equiv|\langle a(t)\rangle|=\prod_n |\langle a_n(t)\rangle|,
\end{align}
and similarly $\langle |a(t)|^2\rangle=\prod_n \langle |a_n(t)|^2\rangle$.

\subsection{Spectral diffusion}
\label{sec: Spectral diffusion}
Various protocols measuring qubit decoherence are sensitive to noise at different frequencies. Such protocols measure the occupation probability of a specific quantum state (typically the ground state $\ket{0}$) as a function of time between the protocol pulses. Each experimental data point is obtained by averaging over $N_{\rm runs}$, where each run involves a free evolution period $t$. Some protocols (e.g., Ramsey oscillations) are sensitive to low-frequency noise, i.e.\ to quantities that fluctuate between different runs (on timescale $N_{\rm runs} t$), and some protocols (e.g., various dynamical decoupling schems such as the Hahn spin-echo) are sensitive only to noise at higher frequencies, i.e.\ to quantities that fluctuate within a single run (on timescale $t$).

Motivated by this observation, we define
\begin{align}
	\label{eq:EnergyShift}
	\varepsilon_n(t)=\varepsilon_{n,0}+x_n+y_n(t),
\end{align}
where $\varepsilon_{n,0}$ is the TLS energy splitting in the absence of interaction with thermal TLSs, $x_n$ is the "quasi-static" shift at time $t=0$ due to interaction with thermal TLSs being in a specific configuration, and $y_n(t)$ is the dynamical shift due to flipping of thermal TLSs (note that $y_{n}(t=0)=0$). The initial shift $x_n$ changes from run to run of the experiment but does not vary during a single run. For a classical Markov process, the average $\langle \rangle$ over the realizations of the stochastic process $\varepsilon_{n}(t)$ is uniquely determined by the probability distribution function $P(x_n)$ for the initial state $x_n$ of the process and the conditional probability distribution function $P(y_n+x_n;t|x_n;0)$ for the state at time $t$ given the initial state.

According to Klauder and Anderson~\cite{KlauderAndersson62}, at short times where $|y_n(t)|$ is much smaller than the ultimate span of the diffusion (this is introduced below and is called $\mu_{{\rm av},n}$), the probability distribution $P(y_n+x_n;t|x_n;0)$ depends only on $y_n$ and has the form
\begin{align}
	\label{eq:eq6}
	P(y_n+x_n;t|x_n;0)=\frac{1}{2\pi}\int^{\infty}_{-\infty}e^{i\tau y_n-tf_n(\tau)}d\tau\ .
\end{align}
Here the function $f_n(\tau)$ characterizes the diffusion process. One gets Gaussian diffusion if $f_n(\tau)\propto\tau^2$ and Lorentzian diffusion if $f_n(\tau)\propto|\tau|$. We consider the more detailed model of Klauder and Anderson~\cite{KlauderAndersson62}, in which fluctuations in $\varepsilon_n$ are due to the coupling of quantum TLSs to thermal TLSs~\cite{BlackHalperin77}, which fluctuate between two states with the typical switching rate $\kappa=T_{1,\rm{T}}^{-1}$. Within an ensemble of thermal TLSs the switching rate varies over a wide range~\cite{BlackHalperin77}. Yet, for the observables of interest to us, taking $\kappa$ to be the typical switching rate preserves their qualitative behavior, and is assumed henceforth. The coupling is assumed to be dipolar, decaying as $1/r^{3}$ ($r$ is the distance between the quantum TLS and the thermal TLS), with two possible values $\pm v_n/r^{3}$ depending on the state of the thermal TLS. The function $f_n(\tau)$ in this case is found to be $\propto |\tau|$ for $|\tau|>\rho_n\equiv\pi r^{3}_{\rm{min}}/(2v_n)$ ($r^{}_{\rm{min}}$ is the short-distance cutoff, such that $v_n/r^{3}_{\rm{min}}$ is the largest possible coupling) and $\propto\tau^2$ for $|\tau|<\rho_n$. These two limiting cases can be interpolated by the function $f_n(\tau)=m_n\left(\sqrt{\tau^2+\rho^{2}_n}-\rho_n\right)$, where $m_n=(2/3)\pi^{2}v_{n}{\cal N}\kappa\approx(v_n/\bar{r}^{3})\kappa$, with $\bar{r}$ being the average distance between the thermal TLSs (their density is ${\cal N}\approx 1/\bar{r}^{3}$). The functional form of $P(y_n+x_n;t|x_n;0)$ as a function of $y_n$ is Lorentzian at short times $t \ll 1/(m_n\rho_n)$. In this regime the essential decay of the exponential function inside the integral in (\ref{eq:eq6}) happens at $\tau \gg \rho_n$, so that $f_n(\tau) \approx m_n |\tau|$. At longer times, $t \gg 1/(m_n\rho_n)$, essential decay happens already at $\tau \ll \rho_n$, so that $f_n \propto \tau^2$, and the functional form of $P(y_n+x_n;t|x_n;0)$ is Gaussian.

It should be emphasized that the distribution function Eq.~(\ref{eq:eq6}) does not represent a stationary distribution since it does not approach a limiting distribution for $t\rightarrow\infty$. As discussed by Klauder and Anderson~\cite{KlauderAndersson62}, the generalization of the distribution function to longer times has the form
\begin{widetext}
	\begin{align}
		\label{eq:eq7}
		&P(y_n+x_n;t|x_n;0)=\frac{1}{2\pi}\int^{\infty}_{-\infty}d\tau e^{i\left[y_n+x_n\left(1-e^{-R_{n}t}\right)\right]\tau}e^{-\int^{t}_{0}dt'f_n(\tau e^{-R_{n} t'})}.
	\end{align}
\end{widetext}
The form of Eq.~(\ref{eq:eq7}) guarantees that this probability distribution has a stationary form at times $t\gg 1/R_{n}$, corresponding to the time scale for which most of the thermal TLSs have flipped at least once. The distribution function in this limit depends only on the final position $x_n+y_n$ irrespective of the initial position. This stationary limit is used below to average over the "quasi-static" shift $x_n$ between different runs of a dephasing protocol (e.g., Ramsey or spin-echo), i.e.\ we define $P_{\infty}(x_n)=P(x_n;t\rightarrow\infty|x_n;0)$, which can be written approximately as
\begin{align}
	\label{eq:eq8}
	&P^{}_{\infty}(x_n)\approx\frac{1}{2\pi}\int^{\infty}_{-\infty}d\tau e^{ix_n\tau}e^{-f_n(\tau)/R_{n}}.
\end{align}

We take $R_{n}\equiv R$ to be the typical switching rate  of the thermal TLSs defined above $R=\kappa =T_{1,\rm{T}}^{-1}$. We also define $\mu_{\rm{av},n}=v_n/\bar{r}^{3}$ and $\mu_{\rm{max},n}=v_n/r^{3}_{\rm{min}}$. Then $m_n$ defined above is $\mu_{\rm{av},n}R$ and $\rho_n=1/\mu_{\rm{max},n}$. We therefore obtain
\begin{align}
	\label{eq:eq9}
	&P_{\infty}(x_n)\approx\frac{1}{2\pi}\int^{\infty}_{-\infty}d\tau e^{ix_n\tau}e^{-\mu^{}_{\rm{av},n}\left(\sqrt{\tau^{2}+\rho^{2}_n}-\rho_n\right)}.
\end{align}
Assuming $\mu_{\rm{av},n}\ll\mu_{\rm{max},n}$, the distribution function $P_{\infty}(x_n)$ is has the Lorentzian form $P_{\infty}(x_n)=(\mu_{\rm{av},n}/\pi)/(\mu^{2}_{\rm{av},n}+x^{2}_{n})$ for $|x_n|<\mu_{\rm{max},n}$, and should decay faster than $1/x^{3}_{n}$ for $|x^{}_{n}|>\mu_{\rm{max},n}$, such that the second moment $\braket{x^{2}_{n}}$ is finite.

\section{Analysis}
\label{sec: analysis}
\subsection{Expansion at short times}
\label{sec: short-time expansion}
To further investigate Eq.~(\ref{eq:Dephasing}), we limit ourselves to the short-time expansion of Eq.~(\ref{eq:eq5}), up to the second order in $C_n$. This approximation may be justified by two quite different arguments:

1) In many cases multiple resonant TLSs contribute simultaneously. Then, even if each individual $a_n(t)$ has only slightly deviated from $1$ and can be expanded, the product $a(t) = \prod_n a_n(t)$ has already substantially decayed. This consideration is a variation of the central limit theorem in statistics.

2) The usual definition of the dephasing time as the time at which $|\braket{a(t)}|$ (averaging over the realizations of the spectral diffusion) has decayed by a factor of $1/e$ is, nowadays, of only indicative importance in the field of quantum computing. Much more important is the time at which, e.g., $|\braket{a(t)}|$ reaches an error threshold allowed by the requirements of, e.g., quantum error correction. Thus, we might be more interested to calculate the time at which $|\braket{a(t)}| \sim 1-\epsilon$ for some $\epsilon \ll 1$. To answer this question the short-time expansion is usually sufficient.

Thus we expand
\begin{widetext}
\begin{align}
	\label{eq:anExpansion}
	a_n(t)=e^{-\int^{t}_{0}C_n(t')dt'}\approx 1-\int^{t}_{0}\,C_n(t_1)dt_1+\frac{1}{2}\int^{t}_{0}\int^{t}_{0}\,C_n(t_1)C_n(t_2)dt_{1}dt_2 + \dots
\end{align}
\end{widetext}
and perform the averaging described in Appendix~\ref{Sec:PureDephasingFromCCorrelator}. For the dephasing of the off-diagonal elements of the density matrix we obtain
\begin{align}
	\label{eq:T2Decay}
	|\langle a(t)\rangle| = \sqrt{\braket{|a(t)|^2}} \cdot D(t)\ ,
\end{align}
where
\begin{align}
	\label{eq:pure_dephasing}
	D(t) \equiv \exp{\left[-\frac{1}{2}\sum_n \int^{t}_{0}\int^{t}_{0}\braket{\delta C_n(t_1)\delta C^{\ast}_n(t_2)}dt_1dt_2\right]}
\end{align}
is the pure dephasing decay factor. In the simplest case one would expect $\sqrt{\braket{|a(t)|^2}} \sim \exp[-\Gamma_1 t/2]$, $|\langle a(t)\rangle|=\exp[-\Gamma_2 t]$, $D(t) \sim \exp[-\Gamma_\varphi t]$. Thus, Eq.~(\ref{eq:T2Decay}) stands in clear correspondence to the well-known Bloch's relation $\Gamma_2 = \Gamma_1/2 + \Gamma_\varphi$. Of course, in reality, these decay laws are not necessarily exponential. The quantity $D(t)$ is the central subject of our study.

\subsection{The correlators}

Next we employ the spectral diffusion propagators~(\ref{eq:eq6}) and (\ref{eq:eq9}).
Using Eqs.~(\ref{eq:eq4}) and~(\ref{eq:EnergyShift}), the first moment $\braket{C_n(t)}$ can be written as
\begin{widetext}
\begin{align}
	\label{eq:eq19}
	\braket{C_n(t)}=\frac{g^{2}_{n}}{4}\sum^{\infty}_{m=-\infty}J^{2}_{m}\int^{\infty}_{-\infty}dx_n P_{\infty}(x_n)\int^{\infty}_{-\infty}\frac{P(y_n+x_n;t|x_n;0)}{\gamma_n-i(\Delta_n-x_n-y_n+m\Omega)}dy_n,
\end{align}
where $\Delta_n\equiv E_0-\varepsilon_{n,0}$. Here we have averaged over the spectral diffusion realizations $x_n$ and $y_n$ using the distribution functions $P_{\infty}(x_n)$ and $P(y_n+x_n;t|x_n;0)$ discussed above. The correlator $\braket{C_n(t_1)C^{\ast}_n(t_2)}$ can be written similarly as
\begin{align}
	\label{eq:eq20}
	\braket{C_n(t_1)C^{\ast}_n(t_2)}=\,&\frac{g^{4}_{n}}{16}\sum^{\infty}_{m,m'=-\infty}J^{2}_{m}J^{2}_{m'}\int^{\infty}_{-\infty}dx_n P_{\infty}(x_n)\int^{\infty}_{-\infty}\frac{P(y_{n2}+x_n;t_2|x_n;0)}{\gamma_n+i(\Delta_n-x_n-y_{n2}+m\Omega)}dy_{n2}\nonumber\\
	&\times\int^{\infty}_{-\infty}\frac{P(y_{n1}+y_{n2}+x_n;t_1|y_{n2}+x_n;t_2)}{\gamma_n-i(\Delta_n-x_n-y_{n1}-y_{n2}+m'\Omega)}dy_{n1},
\end{align}
where we assumed $t_1>t_2$.
\end{widetext}
\subsection{Estimation of the pure dephasing factor $D(t)$}

The estimation of $D(t)$, defined in Eq.~(\ref{eq:pure_dephasing}), is performed in Appendix~\ref{Sec:EstimatingDt} separately for $t\ll 1/R\equiv T_{1,\rm{T}}$ and for $t \gg T_{1,\rm{T}}$. This is because the spectral diffusion undergoes a qualitative change at times of order $T_{1,\rm{T}}$ as described in Sec.~\ref{sec: Spectral diffusion}. At $t\ll T_{1,\rm{T}}$ the distribution function $P(y_n+x_n;t|x_n;0) \sim \frac{W_n(t)}{W^{2}_n(t)+y^{2}_{n}}$ is a Lorentzian depending on the relative deviation $y_{n}$ only and the width $W_n(t)$ grows linearly in time (for details see Sec.~\ref{sec: Spectral diffusion} and Appendix~\ref{Sec:EstimatingDt}). At $t \gg T_{1,\rm{T}}$ the distribution function $P(y_n+x_n;t|x_n;0)$ saturates at $P_{\infty}(y_n+x_n)$.

One finally ends up with
\begin{widetext}
\begin{align}
	\label{eq:eq45}
	-2 \ln D(t)
	\sim\frac{g_{\rm max}^{4}}{\delta_{\rm typ}}\left(\sum^{\infty}_{m=-\infty}J^{4}_{m}\right)\begin{cases}
		\frac{1}{\gamma}t^2, \quad t\ll\frac{\gamma}{\mu_{\rm{av}}}T_{1,\rm{T}}\\
		\frac{T_{1,\rm{T}}}{\mu_{\rm{av}}}t\ln\left(\frac{\mu_{\rm{av}}}{T_{1,\rm{T}}\gamma}t\right), \quad T_{1,\rm{T}} \gg t \gg\frac{\gamma}{\mu_{\rm{av}}}T_{1,\rm{T}}\\
		\frac{T_{1,\rm{T}}}{\mu_{\rm{av}}}t\ln\left(\frac{\mu_{\rm{av}}}{\gamma}\right), \quad t\gg T_{1,\rm{T}}\,\, .
	\end{cases}
\end{align}
\end{widetext}
Here, $\gamma$ and $\mu_{\rm av}$ stand for typical values (see Appendix~\ref{Sec:EstimatingDt}).
The case $t\ll  T_{1,\rm{T}}$ is subdivided into the regime of quasi-static dephasing ($t\ll\frac{\gamma}{\mu_{\rm{av}}}T_{1,\rm{T}}$), in which the random phase is dominated by the distribution of the initial spectral positions $x_n$, and the regime of dynamical dephasing, in which the spectral diffusion during the time interval $t$ matters (see more details in Appendix~\ref{Sec:EstimatingDt}). The validity of this result is insured at least as long as $-2 \ln [D(t)] \lesssim 1$ as discussed in Appendix~\ref{Sec:EstimatingDt}. Thus it can be safely used in order to estimate the effective dephasing rate.

\section{Discussion}
\label{sec: discussion}
Here we analyze the pure dephasing law given by Eq.~(\ref{eq:eq45}).
We first start the discussion in the absence of modulations: $A=0$, in which case $\sum^{\infty}_{m=-\infty}J^{4}_{m}=1$ as $J_{m}(0)=\delta_{m,0}$.
First, we notice that the condition (\ref{eq:T1Tcondition}) assumed in the course of the derivation and required for the validity of the Markovian approximation is equivalent to $T_{1,T}\gamma (\gamma/\mu_{\rm av})>1$. This follows from $W_{n}(t)\approx(\mu_{\rm{av},n}/T_{1,\rm{T}})t$ and the requirement that the maximum deviation in $\varepsilon_{n}(t)$ during the time interval of order $1/\gamma_n$ is smaller than $\gamma_n$.

\begin{figure}
	\includegraphics[width=0.5\textwidth,height=7cm]{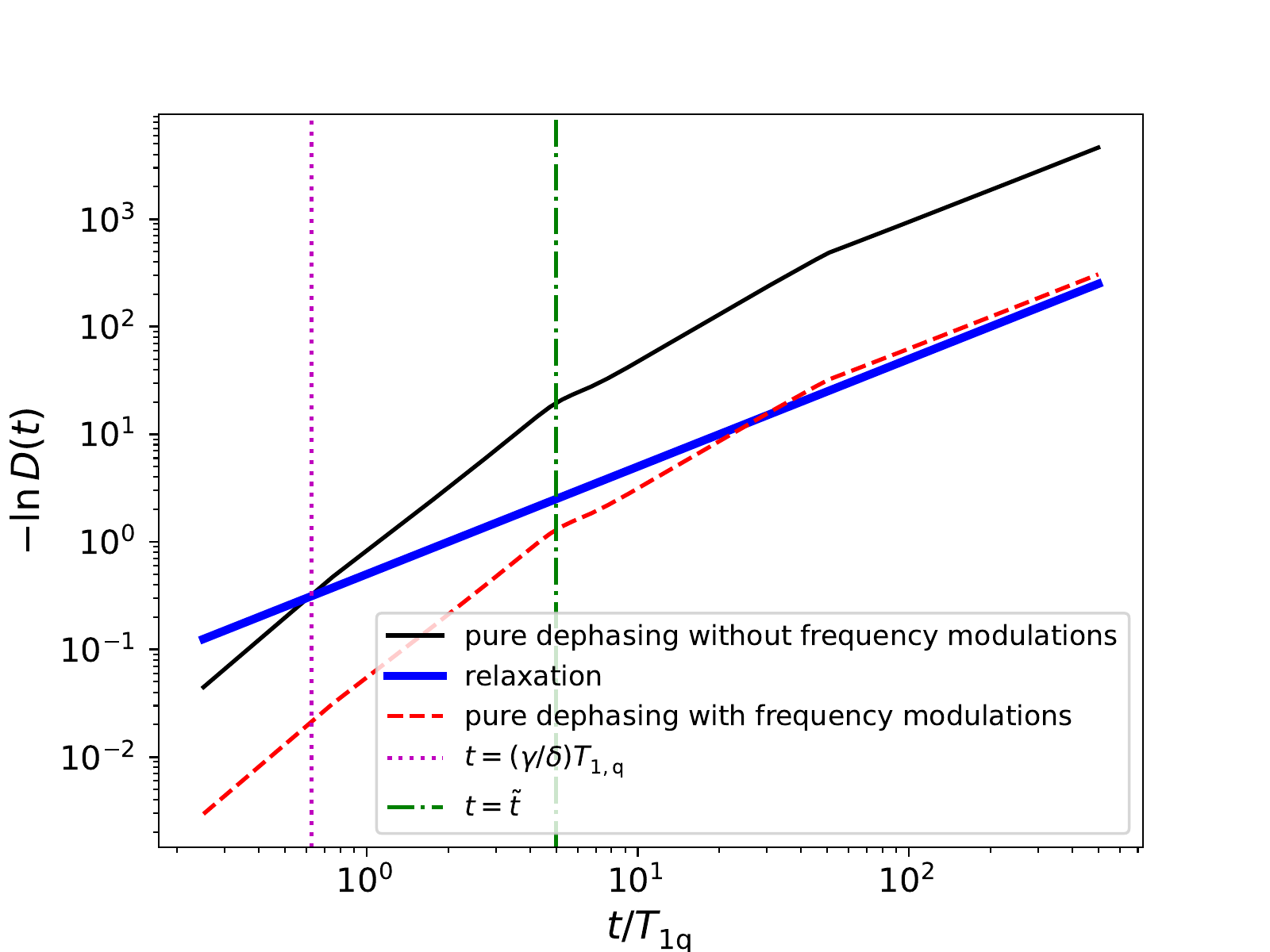}
	\caption{The pure dephasing exponent $-\ln D(t)$ Eq.~(\ref{eq:eq45}) due to spectral diffusion of quantum TLSs as a function of time (in units of the qubit relaxation time $T_{1,\rm{q}} = \Gamma^{-1}_{1,\rm{q}}$) in the regime $\gamma<\mu_{\rm av}$ and $\left(\Gamma_{1,\rm{q}}T_{1,\rm{T}}\right)^2\,\left(\gamma\delta_{\rm typ}/\mu^2_{\rm{av}}\right)\gg 1$ (the specific parameters are $\delta_{\rm typ} / 2\pi = 0.8\,$MHz, $\gamma / 2\pi = 0.5\,$MHz, $\mu_{\rm{av}} / 2\pi = 5\,$MHz, $T_{1,\rm{q}} = \Gamma^{-1}_{1,\rm{q}} = 20\,\mu$s, and $T_{1,\rm{T}} = 1\,$ms~\cite{KlimovPRL18, Lisenfeld2016, Matityahu2016}). In this regime the pure dephasing exponent without frequency modulations (thin solid black) is comparable in magnitude to the relaxation exponent $\Gamma_{1, \mathrm{q}}t/2$ (thick solid blue) at short times $t < \tilde{t}$ [indicated by the dash-dotted green vertical line, see Eq.~(\ref{eq:crossover})], with crossing time $t \approx (\gamma/\delta)\Gamma^{-1}_{1, \mathrm{q}}$ indicated by the dotted magenta vertical line. The red dashed curve is the pure dephasing exponent in the presence of frequency modulations with $A/\Omega = 10$, which suppresses the pure dephasing process and allows to restore the relaxation-limited decoherence rate $\Gamma_{2} = \Gamma_{1,\rm{q}}/2$.}
	\label{fig:dephasing_law}
\end{figure}

To decide which contribution is the relevant one for the actual qubit dephasing, we evaluate the "quasi-static" dephasing at the crossover time
\begin{align}
	\label{eq:crossover}
	\tilde{t}=\frac{\gamma}{\mu_{\rm{av}}}T_{1,\rm{T}}\ .
\end{align}
If it is larger than some threshold $\epsilon$, then the "quasi-static" contribution is the relevant one. Otherwise, the dephasing occurs only at longer times where the dynamic contribution is important. From Eq.~(\ref{eq:eq45}) we find
\begin{align}
	\label{eq:eq39}
	-2 \ln D(\tilde{t}) \sim \frac{g_{\rm max}^{4}}{\delta_{\rm typ}}\frac{\gamma\,T^2_{1,\rm{T}}}{\mu^{2}_{\rm{av}}}\sim\left(\Gamma_{1,\rm{q}}T_{1,\rm{T}}\right)^2\frac{\gamma\delta_{\rm typ}}{\mu^2_{\rm{av}}},
\end{align}
where we have used the Fermi's golden rule estimate for the qubit energy relaxation rate, $\Gamma_{1,\rm{q}}\sim g_{\rm max}^2/\delta_{\rm typ}$ (see, e.g., Ref.~\cite{MatityahuShnirmanSchechter21}). We are working in the regime $\gamma \ll \mu_{\rm{av}}$. It is also natural to assume $\delta_{\rm typ} \sim \gamma \ll \mu_{\rm{av}}$ because in this case the randomly looking ''$T_1$ fingerprint''~\cite{KlimovPRL18,SchloerPRL19,BurnettNPJQI19} is reproduced. Then the number of contributing resonant TLSs $N\sim \mu_{\rm{av}}/\delta_{\rm typ} \log(...)$ is large and the short-time expansion for each $n$ is sufficient. Let us, first, consider the limit $-2\ln D(\tilde{t}) = \left(\Gamma_{1,\rm{q}}T_{1,\rm{T}}\right)^2\,\left(\gamma\delta_{\rm typ}/\mu^2_{\rm{av}}\right)\gg 1$. That would necessarily (but not sufficiently) require $\Gamma_{1,\rm{q}}T_{1,\rm{T}} \gg 1$, i.e., the thermal TLSs relax slower than the qubit.  Since $\Gamma_{1, \mathrm{q}}$ is typically smaller than $\gamma$, the condition $\left(\Gamma_{1,\rm{q}}T_{1,\rm{T}}\right)^2\,\left(\gamma\delta_{\rm typ}/\mu^2_{\rm{av}}\right) \gg1$ is more restrictive than the Markovian condition $T_{1,T}\gamma (\gamma/\mu_{\rm av})>1$ discussed above. Thus, the validity of the Markovian condition~(\ref{eq:T1Tcondition}) can be safely assumed. In this regime the substantial dephasing happens already during the period dominated by the initial (quasi-static) fluctuations, i.e., $-2 \ln D(t) \sim g_{\rm max}^4 t^2/(\gamma\delta_{\rm typ}) \sim \left(\Gamma_{1,\rm{q}}\,t\right)^2 (\delta_{\rm typ}/\gamma)$. Thus, depending on the ratio $\delta_{\rm typ}/\gamma$ the pure dephasing rate can become comparable or even somewhat larger than the relaxation rate  $\Gamma_{1,\rm{q}}$. The dephasing law Eq.~(\ref{eq:eq45}) in this regime is sketched in Fig.~\ref{fig:dephasing_law}, where we compare the behavior of the pure dephasing exponent $-\ln D(t)$ (thin solid black curve) with the relaxation exponent $\Gamma_{1, \mathrm{q}}t/2$ (thick solid blue curve) for realistic parameters for modern superconducting qubits~\cite{KlimovPRL18, Lisenfeld2016, Matityahu2016}. The pure dephasing exponent is comparable to the relaxation exponent (they become equal at $t \approx (\gamma/\delta)\Gamma^{-1}_{1, \mathrm{q}}$ as indicated by the dotted magenta vertical line) in the quasi-static regime $t < \tilde{t}$ where the dephasing process occurs ($\tilde{t}$ is indicated by a dash-dotted green vertical line).

In the other limit $\left(\Gamma_{1,\rm{q}}T_{1,\rm{T}}\right)^2\,\left(\gamma\delta_{\rm typ}/\mu^2_{\rm{av}}\right)\lesssim 1$ (but still $T_{1,T}\gamma (\gamma/\mu_{\rm av})>1$), the dephasing happens at longer times, $t\gg\tilde{t}$. In this case, with logarithmic accuracy, the pure dephasing rate can be estimated to be $\sim \frac{g_{\rm max}^{4}}{\delta_{\rm typ}}\,\frac{T_{1,\rm{T}}}{\mu_{\rm{av}}} \sim \Gamma_{1,\rm{q}}\left(\Gamma_{1,\rm{q}}\,T_{1,\rm{T}}\right)\left(\delta_{\rm typ} /\mu_{\rm{av}}\right)$. Thus it is necessarily much smaller than $\Gamma_{1,\rm{q}}$ and the pure dephasing due to the spectral diffusion is subdominant. Both limits could be relevant depending on the qubit relaxation rate and the relaxation rate of the thermal TLSs, the latter being strongly temperature dependent.

Our results are obtained for the limit where $\mu_{\rm av}>\gamma$. In the opposite limit, i.e. for $\mu_{\rm av}<\gamma$, pure dephasing is suppressed and dominated by the quasi-static contribution. In the ultimate regime of $\mu_{\rm max}< \gamma$ the effective dephasing rate is smaller by a factor of $\mu_{\rm max}/\gamma$ compared to its value in the limit of $\mu_{\rm av}>\gamma$, as given in Eq.~(\ref{eq:eq45}) (see Appendix~\ref{Sec:SmallDiffusionLimit} for derivation of pure dephasing in the limit of $\mu_{\rm max}<\gamma$). 

Our results reflect the dephasing process, in which maximal dephasing of the qubit is related to a shift of a resonant TLS by an energy which is of the order of its width $\gamma$. For $\mu_{\rm av}<\gamma$ such processes are not allowed, and thus qubit dephasing is small. For $\mu_{\rm av}>\gamma$ quasistatic dephasing dominates for $t<\tilde{t}$, which marks the border of the $t^2$ behavior of the dephasing. At longer times dynamical processes shift the TLS energies randomly, typically by more than their width $\gamma$. The dephasing process becomes a dynamical one and the effective dephasing rate is reduced as compared with the static regime [cf. Eq.~(\ref{eq:eq45})]. 

The dependence of the pure dephasing rate on $\mu_{\rm av}$ and $T_{1,\rm{T}}$ dictates its dependence on temperature. For very low temperatures $\mu_{\rm av}<\gamma$, and pure dephasing is weak. Yet, once the temperature is large enough so that $\mu_{\rm av}>\gamma$, we find an intriguing decrease in pure dephasing with increased temperature. This decrease is a consequence of the diminishing of the regime  $t<\gamma\, T_{1,\rm{T}}/\mu_{\rm av}$ in which dephasing is quasi-static, as $T^{-1}_{1,\rm{T}} \propto T^3$ (Refs.~\cite{AndersonHalperinWarma72,Phillips1972,BlackHalperin77}), and $\mu_{\rm av} \propto T$ (Ref.~\cite{BlackHalperin77}). With increased temperature, dynamical effects which slow the growth of dephasing enter at earlier times, causing relative reduction in dephasing rate which reaches a factor of $1/T^4$ at long times. As a result, we find a reentrant temperature dependence of the pure dephasing rate, which peaks at the temperature where $\mu_{\rm av}=\gamma$. 

Let us now discuss the effect of qubit frequency modulations on the pure dephasing decay. We observe from Eq.~(\ref{eq:eq45}) that the factor $\sum^{\infty}_{m=-\infty}J^{4}_{m}\sim\Omega/A$ reduces effectively the dephasing rate for $\Omega \ll A$ (the fact that $\sum^{\infty}_{m=-\infty}J^{4}_{m}(A/\Omega)\sim\Omega/A$ results from the typical value $|J^{}_{m}(x)|\sim\sqrt{\Omega/A}$ of the Bessel functions for $|m|<x$, see Ref.~\cite{MatityahuShnirmanSchechter21}). From Eq.~(\ref{eq:eq45}) one sees that the effective rate is reduced by a factor of $\sqrt{\Omega/A}$ in the case that the quasi-static contribution dominates pure dephasing, i.e., for $t\ll\tilde{t}$. For longer times, when the dynamical contribution dominates, the modulations reduce the pure dephasing rate by a factor $\sim \Omega/A$. In addition, the estimate of the dephasing at the crossover between the static and the dynamic regimes (\ref{eq:eq39}) is also modified and reads now
\begin{align}
	\label{eq:eq39withJm}
	-2 \ln D(\tilde{t}) &\sim \frac{g_{\rm max}^{4}}{\delta_{\rm typ}}\left(\sum^{\infty}_{m=-\infty}J^{4}_{m}\right)\frac{\gamma\,T^2_{1,\rm{T}}}{\mu^{2}_{\rm{av}}}\nonumber\\
	&\sim\left(\Gamma_{1,\rm{q}}T_{1,\rm{T}}\right)^2\left(\sum^{\infty}_{m=-\infty}J^{4}_{m}\right)\frac{\gamma\delta_{\rm typ}}{\mu^2_{\rm{av}}}\ .
\end{align}
The quasi-static contribution at $t=\tilde t$ is reduced by the modulations, thus shrinking the parameter domain where its contribution dominates. On the other hand, in the dynamical regime, $t>\tilde t$, pure dephasing is subdominant even in the absence of modulations. The reduction of the pure dephasing exponent Eq.~(\ref{eq:eq45}) due to frequency oscillations with $A/\Omega = 10$ is illustrated by the dashed red curve in Fig.~\ref{fig:dephasing_law}.

We note that in contrast to our previous analysis of the effect of the modulations on the qubit relaxation rate~\cite{MatityahuShnirmanSchechter21}, in which the modulations did not affect the average relaxation rate but only reduced its temporal fluctuations, here we obtain a reduction of the average pure dephasing rate. 
Such reduction of pure dephasing is certainly significant in the regime where pure dephasing rate is comparable of larger than relaxation induced dephasing rate $\Gamma_{1,\rm{q}}/2$. However, given recent results showing that relaxation induced dephasing can be mitigated algorithmically~\cite{Kubica2023,Levine2023}, mitigation of pure dephasing, as shown here, is of utmost significance also in the regime where relaxation induced dephasing dominates.
Similar results of stabilisation of relaxation time and reduction of pure dephasing rate were recently obtained by engineering the TLS noise spectral densities~\cite{You2022}.

In this paper we considered the pure dephasing process defined by the Ramsey protocol. We have observed that  periodic modulations can mitigate this dephasing. In the regime where the pure dephasing dominates over relaxation induced dephasing, it is also quasi-static. Thus, it can be mitigated by the echo technique as well. 
Remarkably, periodic modulations mitigate the non-static (exponential) dephasing as well, whereas the echo technique would 
be inefficient in this regime. Although the non-static dephasing is always subdominant, its suppression is very important in view of the ideas of Refs.~\cite{Kubica2023,Levine2023}.

\section{Conclusions}
\label{sec: conclusions}

In this paper we have analysed the pure dephasing of a qubit due to the spectral diffusion of a quantum TLS, which in the first place are responsible for the energy relaxation time of the qubit, $T_{1,\rm{q}}$. We have identified a regime, in which the contribution of the spectral diffusion to the pure dephasing rate is comparable to $T_{1,\rm{q}}^{{-1}}$. This leads to an appreciable deviation from the relation $T_{2,\rm{q}} = 2T_{1,\rm{q}}$,  as is indeed the case in numerous experiments (see, e.g., Refs.~[\onlinecite{PaikPRL2011,SchloerPRL19,BurnettNPJQI19,KandalaNature2019}]). 

This result should be compared to the alternative mechanisms that could violate the relation $T_{2,\rm{q}} = 2T_{1,\rm{q}}$. We first mention that the direct coupling of the thermal fluctuators to the transmon qubit does not contribute to the pure dephasing since transmon qubits are exponentially decoupled from the low frequency charge noise~\cite{Koch2007}. On the other hand, the fluctuations of the critical current~\cite{VanHarlingen2004,Nugroho2013} or the noise of  the photon occupation number in the nearby readout resonator~\cite{Gambetta2006} could be relevant. The latter mechanism can be suppressed in devices with tunable coupling~\cite{Zhang2017}.

To distinguish our pure dephasing mechanism from that caused by the photon shot noise, we note that the latter is characterised by more or less white noise spectrum. Thus the pure dephasing  due to the photon shot noise is Markovian and the time decay is strictly exponential. In contrast, in our case the dominant part of the time decay is that given by the quasi-static noise, thus the time decay takes roughly the form $e^{-\Gamma_{\varphi}^{2}t^{2}}$. 

On the other hand, the pure dephasing due to the critical current noise is caused by thermal TLSs, thus giving a similar quasi-static time decay. Distinguishing the pure dephasing resulting from critical current fluctuations from the mechanism discussed in this paper would be possible via a response to an external electric field. The effect of such field on the TLSs causing the critical current fluctuations would be significantly weaker, since these are located in the Josephson junction itself~\cite{Lisenfeld2019}.

We further analyse the effect of the spectral diffusion in the presence of periodic modulations of the qubit frequency and conclude that in this regime the modulations can strongly suppress the pure dephasing and, ultimately, restore the relation $T_{2,\rm{q}} = 2T_{1,\rm{q}}$, therefore enhancing the performance of superconducting qubits.

Recently it was suggested that relaxation induced dephasing can be overcome in a certain type of error-correction schemes 
(erasure errors) using ''dual rail'' qubits~\cite{Kubica2023,Levine2023}. This leaves pure dephasing as the bottleneck for quantum computation even in the regime where it is subdominant. Thus, mitigation of pure dephasing by periodic modulations as suggested here can be of considerable value also when pure dephasing is weak.

Analysis of periodic bias modulation in this work considers the application of the periodic biasing to the qubit's frequency. In the same way, the periodic bias field can be applied to modulate the excitation energies of the TLSs \cite{KhalilPRB2014,RosenPRL2016,MatityahunpjQuantum2019,YuSciRep2022}. In this case detailed analysis of the dependence of the modulation on individual TLS parameters has to be performed. Yet, we expect the results presented in this paper to be applicable, up to quantitative corrections, to the case where the periodic bias field is applied to the TLSs.

The phenomenon of pure dephasing considered in this paper is related to the problem of frequency noise in resonators \cite{BurnettNatComm2014,FaoroIoffePRB2015,BurinPRB2015}. In both systems thermally excited TLSs interact with resonant TLSs, and the energy shifts of the latter cause, via the Lamb shift, fluctuations in the resonance energy of the resonator, or in the excitation energy of the qubit. Theory for frequency noise in resonators requires the application of spectral diffusion theory \cite{KlauderAndersson62,BlackHalperin77} to this specific problem. This was done directly by Burin et. al.~\cite{BurinPRB2015}, and via separation of the different contributions of strongly and weakly interacting thermal TLSs in the work of Faoro and Ioffe~\cite{FaoroIoffePRB2015}(see~\footnote{The two approaches are in principle equivalent, the requirement in Ref.~[\onlinecite{FaoroIoffePRB2015}] for an energy dependent DOS is consequence of their neglect of the logarithmic expansion in time of the spectral diffusion width and its temperature dependence.}). Here we follow an approach similar to that of Burin et. al.~\cite{BurinPRB2015}, directly applying the theory of Klauder and Anderson~\cite{KlauderAndersson62} and treating all thermal TLSs on the same footing. Yet, in similarity to Faoro and Ioffe~\cite{FaoroIoffePRB2015}, we are able to point to the source of strong pure dephasing being the spectral diffusion  over energy shifts of the resonant TLSs of order of their resonance width, which corresponds to "strong interactions" in Ref.~[\onlinecite{FaoroIoffePRB2015}]. Reentrant temperature dependence, as is found for the frequency noise in resonators~\cite{FaoroIoffePRB2015,BurinPRB2015}, is also found here for the pure dephasing. This effect is stronger here, as temperature dictates not only the width of spectral diffusion, but also the time scale for thermal TLS relaxation, which is dominant in dictating pure dephasing.

\section*{Acknowledgments}
M.S. acknowledges support by the Israel Science Foundation (Grant No. 2300/19). A.S. was supported by the Baden-Württemberg-Stiftung (Project QuMaS) and by the German Ministry of Education and Research (BMBF) within the project QSolid (FKZ: 13N16151).

\appendix

\section{Derivation of the non-Hermitian Hamiltonian}
\label{Sec:NHHamiltonian}
The full Hamiltonian of our system reads $\mathcal{H}^{}_{}(t)=\mathcal{H}^{}_{\mathrm{Qubit-TLS}}(t)+\mathcal{H}^{}_{\mathrm{TLS-Phonons}}+\mathcal{H}^{}_{\mathrm{Phonons}}$, where $\mathcal{H}^{}_{\mathrm{Qubit-TLS}}(t)$ is given by Eq.~(1) of the main text and
\begin{align}
	\label{eq:1B}&\mathcal{H}^{}_{\mathrm{TLS-Phonons}}=\frac{1}{2}\sum^{}_{n,k}\tau^{(n)}_{x}(v^{}_{nk}a^{}_{k}+v^{\ast}_{nk}a^{\dag}_{k})\ ,\nonumber\\
	&\mathcal{H}^{}_{\mathrm{Phonons}}=\sum^{}_{k}\omega^{}_{k}a^{\dag}_{k}a^{}_{k}\ .
\end{align}
Here $a^{\dag}_{k}$ and $a^{}_{k}$ are the phonon creation and annihilation operators, $\omega^{}_{k}$ are the phonon frequencies, and $v^{}_{nk}$ describes the coupling between the $n$th TLS and the phonon mode with wave vector $k$. We now consider the interaction picture with respect to the perturbation $V=(1/2)\sum^{}_{n}g^{}_{n}\sigma^{}_{x}\tau^{(n)}_{x}+(1/2)\sum^{}_{n,k}v^{}_{nk}\tau^{(n)}_{x}(a^{}_{k}+a^{\dag}_{k})$ and employ the rotating wave approximation. Within the single excitation subspace spanned by the states $\ket{1}\ket{\{0\}^{}_{\mathrm{TLS}}}\ket{\{0\}^{}_{\mathrm{Phonons}}}$, $\ket{0}\ket{\{1_{n}\}^{}_{\mathrm{TLS}}}\ket{\{0\}^{}_{\mathrm{Phonons}}}$ and $\ket{0}\ket{\{0\}^{}_{\mathrm{TLS}}}\ket{\{1_{k}\}^{}_{\mathrm{Phonons}}}$, in which either the qubit, a single TLS, or a single phonon is excited (we use the notation $\ket{\{1_{n}\}^{}_{\mathrm{TLS}}}=\ket{0,\ldots,0,1_{n},0,\ldots,0}$ for the state in which the $n$th TLS is excited and all other TLSs are at their ground states and $\ket{\{1_{k}\}^{}_{\mathrm{Phonons}}}=\ket{0,\ldots,0,1_{k},0,\ldots,0}$ for the state with a single phonon occupying the mode with wave vector $k$). Consider the general state $\ket{\psi^{}_{\mathrm{I}}(t)}=a(t)\ket{1}\ket{\{0\}^{}_{\mathrm{TLS}}}\ket{\{0\}^{}_{\mathrm{Phonons}}}+\sum^{}_{n}b^{}_{n}(t)\ket{0}\ket{\{1_{n}\}^{}_{\mathrm{TLS}}}\ket{\{0\}^{}_{\mathrm{Phonons}}}+\sum^{}_{k}c^{}_{k}(t)\ket{0}\ket{\{0\}^{}_{\mathrm{TLS}}}\ket{\{1_{k}\}^{}_{\mathrm{Phonons}}}$. Note that the qubit reduced density matrix Eq.~(\ref{eq:densitymatrix}) is obtained by tracing over the TLS and phonon states, and using the normalization condition $|a(t)|^2+\sum_{n}|b_n(t)|^2+\sum_{k}|c_k(t)|^2=1$. The Schr\"{o}dinger equation for $\ket{\psi^{}_{\mathrm{I}}(t)}$ reads
\begin{align}
	\label{eq:2B}
	&\dot{a}(t)=-\frac{i}{2}\sum^{}_{n}g^{}_{n}e^{i\left(E^{}_{0}t+\phi(t)-\varepsilon^{}_{n}t\right)}b^{}_{n}(t),\nonumber\\
	&\dot{b}^{}_{n}(t)=-\frac{i}{2}\left[g^{}_{n}e^{-i\left(E^{}_{0}t+\phi(t)-\varepsilon^{}_{n}t\right)}a(t)+\sum^{}_{k}v^{}_{nk}e^{i(\varepsilon^{}_{n}-\omega^{}_{k})t}c^{}_{k}(t)\right],\nonumber\\
	&\dot{c}^{}_{k}(t)=-\frac{i}{2}\sum^{}_{l}v^{\ast}_{lk}e^{-i(\varepsilon^{}_{l}-\omega^{}_{k})t}b^{}_{l}(t),
\end{align}
with initial conditions $a(0)=1$, $b^{}_{n}(0)=c^{}_{k}(0)=0\;\; \forall\, n, k$. Integrating the last equation and substituting into the second, we obtain
\begin{align}
	\label{eq:3B}
	\dot{a}(t)=&-\frac{i}{2}\sum^{}_{n}g^{}_{n}e^{i\left(E^{}_{0}t+\phi(t)-\varepsilon^{}_{n}t\right)}b^{}_{n}(t),\nonumber\\
	\dot{b}^{}_{n}(t)=&-\frac{i}{2}g^{}_{n}e^{-i\left(E^{}_{0}t+\phi(t)-\varepsilon^{}_{n}t\right)}a(t)\nonumber\\
	&-\frac{1}{4}\sum^{}_{l}e^{i(\varepsilon^{}_{n}-\varepsilon^{}_{l})t}\int^{t}_{0}K^{}_{nl}(t-t')b^{}_{l}(t'),
\end{align}
where $K^{}_{nl}(\tau)=\sum^{}_{k}v^{}_{nk}v^{\ast}_{lk}e^{i(\varepsilon^{}_{l}-\omega^{}_{k})\tau}$. We now apply the Weisskopf-Wigner (Markov) approximation~\cite{WeisskopfWigner30}. This is justified because the functions $K^{}_{nl}(\tau)$ decay rapidly on the characteristic time scale $1/\gamma^{}_{l}$ (to be defined below) describing the temporal variation of $b^{}_{l}(t)$. One then replaces $b^{}_{l}(t')$ by $b^{}_{l}(t)$ in the integral and extends the upper limit of the integration to $\infty$ to obtain the following equations:
\begin{align}
	\label{eq:4B}
	\dot{a}(t)=&-\frac{i}{2}\sum^{}_{n}g^{}_{n}e^{i\left(E^{}_{0}t+\phi(t)-\varepsilon^{}_{n}t\right)}b^{}_{n}(t),\nonumber\\
	\dot{b}^{}_{n}(t)=&-\frac{i}{2}g^{}_{n}e^{-i\left(E^{}_{0}t+\phi(t)-\varepsilon^{}_{n}t\right)}a(t)\nonumber\\
	&-\frac{1}{4}\sum^{}_{l}C^{}_{nl}e^{i(\varepsilon^{}_{n}-\varepsilon^{}_{l})t}b^{}_{l}(t),
\end{align}
with $C^{}_{nl}=\int^{\infty}_{0}K^{}_{nl}(\tau)d\tau$. The non-diagonal terms ($n\neq l$) couple different TLSs. These could become important if two or more TLSs are close to degeneracy, $|\varepsilon^{}_{n}-\varepsilon^{}_{l}|\ll \gamma^{}_{l},\gamma^{}_{n}$. Then, cooperative super-radiant and sub-radiant dynamics of this group of TLSs could emerge~\cite{DickePhysRev54,HeppLieb73}. In our case this possibility can be neglected, as the average level spacing $\Delta$ between the TLSs is much larger than the typical width $\gamma^{}_{\mathrm{typ}}$. Even if such a degeneracy happens, but the two TLSs are not spatially close to each other, the phases in $K^{}_{nl}(\tau)=\sum^{}_{k}v^{}_{nk}v^{\ast}_{lk}e^{i(\varepsilon^{}_{l}-\omega^{}_{k})\tau}$ will average this coupling to zero.

We therefore retain only diagonal terms, which gives
\begin{align}
	\label{eq:5B}
	&\dot{a}(t)=-\frac{i}{2}\sum^{}_{n}g^{}_{n}e^{i\left(E^{}_{0}t+\phi(t)-\varepsilon^{}_{n}t\right)}b^{}_{n}(t),\nonumber\\
	&\dot{b}^{}_{n}(t)=-\frac{i}{2}g^{}_{n}e^{-i\left(E^{}_{0}t+\phi(t)-\varepsilon^{}_{n}t\right)}a(t)-\frac{1}{4}C^{}_{nn}b^{}_{n}(t).
\end{align}
Finally, we define $\tilde{b}^{}_{n}(t)=e^{C^{}_{nn}t/4}b^{}_{n}(t)$ and obtain
\begin{align}
	\label{eq:6B}
	&\dot{a}(t)=-\frac{i}{2}\sum^{}_{n}g^{}_{n}e^{i\left[E^{}_{0}t+\phi(t)-(\varepsilon^{}_{n}-iC^{}_{nn}/4)t\right]}\tilde{b}^{}_{n}(t),\nonumber\\
	&\dot{\tilde{b}}^{}_{n}(t)=-\frac{i}{2}g^{}_{n}e^{-i\left[E^{}_{0}t+\phi(t)-(\varepsilon^{}_{n}-iC^{}_{nn}/4)t\right]}a(t).
\end{align}
By absorbing the imaginary part of $C^{}_{nn}$ (Lamb shift) into the definition of $\varepsilon^{}_{n}$, and defining $\gamma^{}_{n}\equiv\operatorname{Re}(C^{}_{nn})/4$, one ends up with Eqs.~(4) of the main text [with $\tilde{b}^{}_{n}(t)$ replaced by $b^{}_{n}(t)$]. By integrating out the phonons, one thus finds an exact description of the system in terms of the non-Hermitian Hamiltonian~(2) of the main text.

\section{Pure dephasing and relaxation in terms of the correlation functions of $C_{n}(t)$}
\label{Sec:PureDephasingFromCCorrelator}

Expanding Eq.~(\ref{eq:eq5}) up to second order in $C_n$ and averaging, we obtain
\begin{widetext}
	\begin{align}
		\label{eq:eq10}
		\braket{a_n(t)}=\braket{e^{-\int^{t}_{0}C_n(t')dt'}}\approx 1-\int^{t}_{0}\braket{C_n(t_1)}dt_1+\frac{1}{2}\int^{t}_{0}\int^{t}_{0}\braket{C_n(t_1)C_n(t_2)}dt_{1}dt_2.
	\end{align}
Denoting $C_n(t)=C'_n(t)+iC''_n(t)$ and taking the absolute value, we obtain up to second order in $C_n(t)$
	\begin{align}
		\label{eq:eq11}
		|\langle a_n(t)\rangle|\approx&\,1-\int^{t}_{0}\braket{C'_n(t_1)}dt_1+\frac{1}{2}\int^{t}_{0}\int^{t}_{0}\left(\braket{C'_n(t_1)C'_n(t_2)}-\braket{C''_n(t_1)C''_n(t_2)}\right)dt_{1}dt_2+\frac{1}{2}\left(\int^{t}_{0}\braket{C''_n(t_1)}dt_1\right)^2\ .
	\end{align}
Similarly, up to the second order in $C_n(t)$
	\begin{align}
		\label{eq:eq13}
		\braket{|a_n(t)|^2}=\braket{e^{-2\int^{t}_{0}C'_n(t')dt'}}\approx&\,1-2\int^{t}_{0}\braket{C'_n(t_1)}dt_1+2\int^{t}_{0}\int^{t}_{0}\braket{C'_n(t_1)C'_n(t_2)}dt_{1}dt_2,
	\end{align}
and therefore
	\begin{align}
		\label{eq:eq14}
		\sqrt{\braket{|a_n(t)|^2}}\approx&\,1-\int^{t}_{0}\braket{C'_n(t_1)}dt_1+\int^{t}_{0}\int^{t}_{0}\braket{C'_n(t_1)C'_n(t_2)}dt_{1}dt_2-\frac{1}{2}\left(\int^{t}_{0}\braket{C'_n(t_1)}dt_1\right)^2.
	\end{align}
As we are interested in the product $\prod_n a_n(t)$, we calculate the logarithms of Eqs.~(\ref{eq:eq11}) and~(\ref{eq:eq14}):
	\begin{align}
		\label{eq:eq15}
		\ln|\langle a_n(t)\rangle|\approx&-\int^{t}_{0}\braket{C'_n(t_1)}dt_1+\frac{1}{2}\int^{t}_{0}\int^{t}_{0}\left(\braket{C'_n(t_1)C'_n(t_2)}-\braket{C''_n(t_1)C''_n(t_2)}\right)dt_{1}dt_2\nonumber\\
		&+\frac{1}{2}\left(\int^{t}_{0}\braket{C''_n(t_1)}dt_1\right)^2-\frac{1}{2}\left(\int^{t}_{0}\braket{C'_n(t_1)}dt_1\right)^2 \nonumber\\
		\ln\sqrt{\braket{|a_n(t)|^2}}\approx&\,-\int^{t}_{0}\braket{C'_n(t_1)}dt_1+\int^{t}_{0}\int^{t}_{0}\braket{C'_n(t_1)C'_n(t_2)}dt_{1}dt_2\nonumber\\
		&-\left(\int^{t}_{0}\braket{C'_n(t_1)}dt_1\right)^2.
	\end{align}
Up to the second order in $C_n(t)$, we can thus write
	\begin{align}
		\label{eq:eq16}
		\ln|\langle a_n(t)\rangle|\approx&\ln\sqrt{\braket{|a_n(t)|^2}}-\frac{1}{2}\int^{t}_{0}\int^{t}_{0}\left(\braket{C'_n(t_1)C'_n(t_2)}+\braket{C''_n(t_1)C''_n(t_2)}\right)dt_{1}dt_2\nonumber\\
		&+\frac{1}{2}\left(\int^{t}_{0}\braket{C'_n(t_1)}dt_1\right)^2+\frac{1}{2}\left(\int^{t}_{0}\braket{C''_n(t_1)}dt_1\right)^2,
	\end{align}
or equivalently,
	\begin{align}
		\label{eq:eq17}
		&\ln\sqrt{\braket{|a_n(t)|^2}}\approx\,-\int^{t}_{0}\braket{C'_n(t_1)}dt_1+\int^{t}_{0}\int^{t}_{0}\braket{\delta C'_n(t_1) \delta C'_n(t_2)}dt_{1}dt_2\ ,
		\\ \label{eq:eq17a}
		&\ln |\langle a_n(t)\rangle|\approx \ln \sqrt{\braket{|a_n(t)|^2}} - \frac{1}{2}\int^{t}_{0}\int^{t}_{0}\braket{\delta C_n(t_1)\delta C_n^{\ast}(t_2)}dt_1dt_2\ ,
	\end{align}
where $\delta C_n(t)\equiv C_n(t)-\braket{C_n(t)}$ and $\delta C'_n(t)\equiv C'_n(t)-\braket{C'_n(t)}$.
	
We are now in a position to perform the product over $n$ and exponentiate. This gives for the energy relaxation ($T_1$):
	\begin{align}
		\label{eq:eq18}
		\sqrt{\braket{|a(t)|^2}}\approx \exp{\left[-\sum_n\int^{t}_{0}\braket{C'_n(t')}dt' +
			\sum_n\int^{t}_{0}\int^{t}_{0}\braket{\delta C'_n(t_1)\delta C'_n(t_2)}dt_1dt_2 \right]}.
	\end{align}
\end{widetext}
The quite unexpected second term in the exponent reflects the effect of the fluctuations. Exponentiating Eq.~(\ref{eq:eq17a}) we obtain Eqs.~(\ref{eq:T2Decay}) and (\ref{eq:pure_dephasing}) of the main text.
	
\section{Estimation of $D(t)$}
\label{Sec:EstimatingDt}

\subsection{The short time limit $t \ll T_{1,\rm{T}}$}

At short times $t\ll T_{1,\rm{T}}$, the relevant distribution function is given by Eq.~(\ref{eq:eq6}). In this regime, the condition $t\ll(m_n\rho_n)^{-1}=(\mu_{\rm{max},n}/\mu_{\rm{av},n})T_{1,\rm{T}}$ holds, and the diffusion is Lorentzian [as argued following Eq.~(\ref{eq:eq6})] with distribution function
\begin{align}
	\label{eq:eq27}
	P(y_n+x_n;t|x_n;0)=\frac{1}{\pi}\frac{W_n(t)}{W^{2}_n(t)+y^{2}_{n}},
\end{align}
which is homogeneous, i.e.\
\begin{align}
	\label{eq:eq28}
	P(y_n+y'_n+x_n;t|y'_n+x_n;t')=\frac{1}{\pi}\frac{W_n(t-t')}{W^{2}_n(t-t')+y^{2}_{n}}.
\end{align}
Here the width $W_n(t)$ is given by~\cite{BlackHalperin77}
\begin{align}
	\label{eq:eq29}
	W_n(t)\propto\mu_{\rm{av},n}\int^{\infty}_{0}\frac{dy}{\cosh^{2}y}\int^{1}_{0}\frac{1-e^{-x^{2}t/T_{1}(y)}}{x}dx,
\end{align}
where $y=E/(2T)$ and $x=\Delta_0/E$. Here $\Delta_0$ and $E$ are the tunneling amplitude and energy splitting of the fluctuating thermal TLSs, respectively. These are broadly distributed according to the standard tunneling model~\cite{Phillips1972,AndersonHalperinWarma72}. In Eq.~(\ref{eq:eq29}) $T_{1}(y)\propto T_{1,\rm{min}}(T)y^{3}\coth y$, with $T_{1,\rm{min}}(T)\propto T^3$ being the relaxation time of the thermal TLSs with the largest value of $\Delta_0$, and thus with the fastest decay rate to the phonon bath. According to Black and Halperin~\cite{BlackHalperin77}, the function $W_n(t)$ is linear in time for $t\ll T_{1,\rm{min}}(T)$ and approaches $\mu_{\rm{av},n}$ logarithmically at times $t\gg T_{1,\rm{min}}(T)$. In what follows we treat thermal TLSs as having a single relaxation time $T_{1,\rm{T}}\equiv 1/R\sim T_{1,\rm{min}}(T)$.

Performing the integrations we obtain
\begin{widetext}
\begin{align}
	\label{eq:eq31}
	&\braket{C_n(t)}=\frac{g^{2}_{n}}{4}\sum^{\infty}_{m=-\infty}J^{2}_{m}\int^{\infty}_{-\infty}\frac{P_{\infty}(x_n)}{\gamma_n+W_n(t)-i(\Delta_n-x_n+m\Omega)}dx_n,\nonumber\\
	&\braket{C_n(t_1)C^{\ast}_n(t_2)}=\frac{g^{4}_{n}}{16}\sum^{\infty}_{m,m'=-\infty}J^{2}_{m}J^{2}_{m'}\frac{2\gamma_n+2W_n(t_2)+W_n(t_1-t_2)+i(m-m')\Omega}{2\gamma_n+W_n(t_1-t_2)+i(m-m')\Omega}\nonumber\\
	&\times\int^{\infty}_{-\infty} \frac{P_{\infty}(x_n)}{\left[\gamma_n+W_n(t_2)+i(\Delta_n-x_n+m\Omega)\right]\left[\gamma_n+W_n(t_2)+W_n(t_1-t_2)-i(\Delta_n-x_n+m'\Omega)\right]}dx_n.
\end{align}
\end{widetext}
We assume $\mu_{\rm{max},n}\gg\mu_{\rm{av},n}$, such that $P_{\infty}(x_n)$ can be taken as a Lorentzian function of width $\mu_{\rm{av},n}$ up to $|x_n|=\mu_{\rm{max},n}$. Since $\mu_{\rm{av},n}\gg W_n(t)$ for $t\ll T_{1,\rm{T}}$, in the regime $\mu_{\rm{max},n}\gg\gamma_n$ the non-lorentzian tails of $P_{\infty}(x_n)$ at $|x_n|>\mu_{\rm{max},n}$ can be neglected, and we obtain
\begin{widetext}
\begin{align}
	\label{eq:eq32}
	&\braket{C_n(t)}\approx\frac{g^{2}_{n}}{4}\sum^{\infty}_{m=-\infty}\frac{J^{2}_{m}}{\mu_{\rm{av},n}+\gamma_n+W_n(t)-i(\Delta_n+m\Omega)},\nonumber\\
	&\braket{C_n(t_1)C^{\ast}_n(t_2)}\approx\frac{g^{4}_{n}}{16}\sum^{\infty}_{m,m'=-\infty}J^{2}_{m}J^{2}_{m'}\frac{2\mu_{\rm{av},n}+2\gamma_n+2W_n(t_2)+W(t_1-t_2)+i(m-m')\Omega}{2\gamma_n+W_n(t_1-t_2)+i(m-m')\Omega}\nonumber\\
	&\frac{1}{\left[\mu_{\rm{av},n}+\gamma_n+W_n(t_2)+i(\Delta_n+m\Omega)\right]\left[\mu_{\rm{av},n}+\gamma_n+W_n(t_2)+W_n(t_1-t_2)-i(\Delta_n+m'\Omega)\right]}.
\end{align}
\end{widetext}
We note that if $\mu_{\rm{av},n}$ and $\mu_{\rm{max},n}$ are comparable, then $P_{\infty}(x_n)$ is not a Lorentzian function, but the results are qualitatively similar. We next assume $\Omega\gg\gamma_n,W_n(t)$ and therefore only terms $m=m'$ contribute to the sum in the second equation of Eqs.~(\ref{eq:eq31}). The results (\ref{eq:eq32}) hold for an arbitrary relation between $\mu_{\rm{av},n}$ and $\gamma_n$. From now on we use the assumed above regime $\mu_{\rm{av},n}\gg\gamma_n, W_n$
(the other regimes are discussed in Appendix~\ref{Sec:SmallDiffusionLimit}). In this regime we otain
\begin{align}
	\label{eq:eq33}
	\braket{C_n(t)}\approx\,&\frac{g^{2}_{n}}{4}\sum^{\infty}_{m=-\infty}\frac{J^{2}_{m}}{\mu_{\rm{av},n}-i(\Delta_n+m\Omega)},\nonumber\\
	\braket{C_n(t_1)C^{\ast}_n(t_2)}\approx\,&\frac{g^{4}_{n}\mu_{\rm{av},n}}{8\left(2\gamma_n+W_n(t_1-t_2)\right)}\nonumber\\
	&\times\sum^{\infty}_{m=-\infty}\frac{J^4_m}{\mu^2_{\rm{av},n}+(\Delta_n+m\Omega)^2}.
\end{align}
Thus, near the resonance ($\Delta_n + m\Omega \approx 0$) we can estimate $|\braket{C_n(t)}|^{2}\propto 1/\mu^{2}_{\rm{av},n}\ll\braket{C_n(t_1)C^{\ast}_n(t_2)}$. Therefore, we obtain $\braket{\delta C_n(t_1)\delta C^{\ast}_n(t_2)}\approx\braket{C_n(t_1)C^{\ast}_n(t_2)}$. For $t\ll T_{1,\rm{T}}$, the diffusion width $W_{n}(t)$ is a linear function of time, $W_{n}(t)\approx(\mu_{\rm{av},n}/T_{1,\rm{T}})t\equiv\alpha_n t$. Therefore, using the relation
\begin{align}
	\label{eq:eq34}
	\int^{t}_{0}\int^{t}_{0}k(t_1-t_2)dt_1dt_2=\int^{2t}_{0}(2t-\tau)k(\tau)d\tau,
\end{align}
we obtain
\begin{widetext}
\begin{align}
	\label{eq:eq35}
	&\int^{t}_{0}\int^{t}_{0}\braket{\delta C_n(t_1)\delta C^{\ast}_n(t_2)}dt_1dt_2\approx\frac{g^{4}_{n}\mu_{\rm{av},n}}{8}\sum^{\infty}_{m=-\infty}\frac{J^{4}_{m}}{\mu^2_{\rm{av},n}+(\Delta_n+m\Omega)^2}\int^{2t}_{0}\frac{2t-\tau}{2\gamma_n+\alpha_n\tau}d\tau\nonumber\\
	&=\frac{g^{4}_{n}\mu_{\rm{av},n}}{4}\left[-\frac{t}{\alpha_n}+\frac{\alpha_n t+\gamma_n}{\alpha^2_n}\ln\left(1+\frac{\alpha_n t}{\gamma_n}\right)\right]\sum^{\infty}_{m=-\infty}\frac{J^{4}_{m}}{\mu^2_{\rm{av},n}+(\Delta_n+m\Omega)^2}.
\end{align}
At times $t\ll\gamma_n/\alpha_n=(\gamma_n/\mu_{\rm{av},n})T_{1,\rm{T}}$, we obtain
\begin{align}
	\label{eq:eq36}
	\int^{t}_{0}\int^{t}_{0}\braket{\delta C_n(t_1)\delta C^{\ast}_n(t_2)}dt_1dt_2\approx\frac{g^{4}_{n}\mu_{\rm{av},n}}{8\gamma_n}\sum^{\infty}_{m=-\infty}\frac{J^{4}_{m}}{\mu^2_{\rm{av},n}+(\Delta_n+m\Omega)^2}\,t^2,
\end{align}
\end{widetext}
which corresponds to the "quasi-static" limit. On the other hand, at longer times $(\gamma_n/\mu_{\rm{av},n})T_{1,\rm{T}}\ll t\ll T_{1,\rm{T}}$, we have
\begin{widetext}
\begin{align}
	\label{eq:eq37}
	\int^{t}_{0}\int^{t}_{0}\braket{\delta C_n(t_1)\delta C^{\ast}_n(t_2)}dt_1dt_2\approx\frac{g^{4}_{n}}{4}T_{1,\rm{T}}\sum^{\infty}_{m=-\infty}\frac{J^{4}_{m}}{\mu^2_{\rm{av},n}+(\Delta_n+m\Omega)^2}\,t\ln\left(\frac{\mu_{\rm{av},n}}{T_{1,\rm{T}}\gamma_n}t\right).
\end{align}
\end{widetext}
To make a rough estimate, we replace all $\mu_{\rm{av},n}$ by a typical value $\mu_{\rm{av}}$ and similarly for $\gamma_n$. Then, for each value of $m$, we sum the contributions of the TLSs that are resonant at this $m$, i.e., such that $\Delta_n+m\Omega \lesssim \mu_{\rm{av}}$. This gives us $d\epsilon = \mu_{\rm{av}}$ in the distribution (\ref{eq:Distributionofg}). Using this distribution we estimate
\begin{align}
\sum_n \frac{g^{4}_{n}}{\mu^2_{\rm{av}}+(\Delta_n+m\Omega)^2} \sim \frac{d\epsilon}{\delta_{\rm typ}\mu_{\rm av}^2}
\int\limits_{0}^{g_{\rm max}} g^3 dg \sim \frac{g_{\rm max}^4}{\delta_{\rm typ}\mu_{\rm av}}\ .
\end{align}
Since our theory holds only for the TLSs weakly coupled to the qubit, i.e., $g_n < \gamma_n$, it would be reasonable to
put $g_{\rm max} \sim \gamma$.
We, thus, arrive at
\begin{widetext}
\begin{align}
	\label{eq:eq38}
	-2 \ln [D(t)] = \sum_n\int^{t}_{0}\int^{t}_{0}\braket{\delta C_n(t_1)\delta C^{\ast}_n(t_2)}dt_1dt_2\sim\frac{g_{\rm max}^{4}}{\delta_{\rm typ}}\left(\sum^{\infty}_{m=-\infty}J^{4}_{m}\right)\begin{cases}
		\frac{1}{\gamma}t^2, \quad t\ll\frac{\gamma}{\mu_{\rm{av}}}T_{1,\rm{T}}\\
		\frac{T_{1,\rm{T}}}{\mu_{\rm{av}}}t\ln\left(\frac{\mu_{\rm{av}}}{T_{1,\rm{T}}\gamma}t\right), \quad  T_{1,\rm{T}} \gg t\gg\frac{\gamma}{\mu_{\rm{av}}}T_{1,\rm{T}}\,\, .
	\end{cases}
\end{align}
\end{widetext}
One might wonder if this result is valid at all relevant times in view of the short-time expansions (\ref{eq:eq10}) and (\ref{eq:eq13}) used in the calculation. Since the number of contributing TLSs is always large $\sim \mu_{\rm{av}}/\delta_{\rm typ}$, the dephasing law (\ref{eq:eq38}) is valid at least as long as $-2 \ln [D(t)] \lesssim 1$. Thus we can safely use (\ref{eq:eq38}) in order to estimate the effective dephasing rate (time). At much longer times
(\ref{eq:eq38}) might be inaccurate.

\subsection{The long time limit $t \gg T_{1,\rm{T}}$}

Next we discuss the long time limit, $t \gg 1/R\equiv T_{1,\rm{T}}$. We approximate
\begin{widetext}
\begin{align}
	\label{eq:eq40}
	\int^{t}_{0}\int^{t}_{0}\braket{\delta C_n(t_1)\delta C^{\ast}_n(t_2)}dt_1dt_2\approx\int^{t}_{T_{1,\rm{T}}}\int^{t}_{T_{1,\rm{T}}}\braket{\delta C_n(t_1)\delta C^{\ast}_n(t_2)}dt_1dt_2,
\end{align}
\end{widetext}
and note that $P(y_n+x_n;t|x_n;0)\rightarrow P_{\infty}(x_n+y_n)$ for $t\gg T_{1,\rm{T}}$ and similarly $P(y_{n1}+y_{n2}+x_n;t_1|y_{n2}+x_n;t_2)\rightarrow P_{\infty}(x_n+y_{n1}+y_{n2})$ for $t_1,t_2\gg T_{1,\rm{T}}$, except for the region $t_1-t_2<T_{1,\rm{T}}$, where Eq.~(\ref{eq:eq28}) holds. Then for $t_1,t_2\gg T_{1,\rm{T}}$, we use
\begin{widetext}
\begin{align}
	\label{eq:eq41}
	\int^{\infty}_{-\infty}\frac{P_{\infty}(x_n)}{\gamma_n-i(\Delta_n-x_n+m\Omega)}dx_n=\frac{1}{\gamma_n+\mu_{\rm{av},n}-i(\Delta_n+m\Omega)},
\end{align}
\end{widetext}
to obtain
\begin{align}
	\label{eq:eq42}
	&\braket{C_n(t)}=\frac{g^{2}_{n}}{4}\sum^{\infty}_{m=-\infty}\frac{J^{2}_{m}}{\gamma_n+\mu_{\rm{av},n}-i(\Delta_n+m\Omega)}.
\end{align}
The second moment is
\begin{widetext}
\begin{align}
	\label{eq:eq43}
	&\braket{C_n(t_1)C^{\ast}_n(t_2)}=\frac{g^{4}_{n}}{16}\sum^{\infty}_{m,m'=-\infty}\frac{J^{2}_{m}J^{2}_{m'}}{\left[\gamma_n+\mu_{\rm{av},n}+i(\Delta_n+m\Omega)\right]\left[\gamma_n+\mu_{\rm{av},n}-i(\Delta_n+m'\Omega)\right]}
\end{align}
for $t_1-t_2\gtrsim T_{1,\rm{T}}$ and
\begin{align}
	\label{eq:eq44}
	&\braket{C_n(t_1)C^{\ast}_n(t_2)}=\frac{g^{4}_{n}}{16}\sum^{\infty}_{m,m'=-\infty}\frac{J^{2}_{m}J^{2}_{m'}}{\left[\gamma_n+\mu_{\rm{av},n}+i(\Delta_n+m\Omega)\right]\left[\gamma_n+W_n(t_1-t_2)-i(\Delta_n+m'\Omega)\right]}
\end{align}
\end{widetext}
for $t_1-t_2\ll T_{1,\rm{T}}$. The contribution to the second integral in the right hand side of Eq.~(\ref{eq:eq40}) vanishes for $t_1-t_2\gtrsim T_{1,\rm{T}}$, whereas for $t_1-t_2\ll T_{1,\rm{T}}$ the integral is the same as that calculated in Eq.~(\ref{eq:eq35}).

One finally ends up with
\begin{widetext}
\begin{align}
	\label{eq:eq45a}
	-2 \ln D(t)
	\sim\frac{g_{\rm max}^{4}}{\delta_{\rm typ}}\left(\sum^{\infty}_{m=-\infty}J^{4}_{m}\right)		\frac{T_{1,\rm{T}}}{\mu_{\rm{av}}}t\ln\left(\frac{\mu_{\rm{av}}}{\gamma}\right), \quad t\gg T_{1,\rm{T}}\,\, .
\end{align}
\end{widetext}
The validity of this result is again insured at least as long as $-2 \ln [D(t)] \lesssim 1$ as discussed right after Eq.~(\ref{eq:eq38}). Thus it can be safely used in order to estimate the effective dephasing rate. Combining Eq.~(\ref{eq:eq38}) and Eq.~(\ref{eq:eq45a}) we obtain Eq.~(\ref{eq:eq45}) of the main text.

\section{The limit \boldmath$\mu_{\rm{av}} \ll \gamma$}
\label{Sec:SmallDiffusionLimit}

As the average span of the diffusion $\mu_{\rm{av},n}$ becomes smaller than the width $\gamma_n$, the results for the
pure dephasing change. Analyzing Eq.~(\ref{eq:eq32}) we observe that the inequality $|\braket{C_n(t)}|^2 \ll \braket{C_n(t_1)C^{\ast}_n(t_2)}$ does not hold anymore. Thus, the fluctuations $\braket{\delta C_n(t_1)\delta C^{\ast}_n(t_2)}$ are suppressed and the dephasing is reduced. In the intermediate regime $\mu_{\rm{av},n} <\gamma_n < \mu_{\rm{max},n}$ the analysis is complicated and we concentrate on the ultimate case $\mu_{\rm{max},n} <\gamma_n$.

In the limit $\mu_{\rm{max},n}\ll\gamma_n$ the center of the Lorentzian functions appearing in Eqs.~(\ref{eq:eq19}) and~(\ref{eq:eq20}) does not fluctuate significantly on the energy scale compared to the width $\gamma_n$ of the Lorentzian. The contributions of the dynamic shifts is thus small and to obtain the contribution of the "quasi-static" dephasing we can take the conditional probabilities to be delta functions of the dynamic shift. We thus end up with
\begin{widetext}
\begin{align}
	\label{eq:eq21}
	&\braket{C_n(t)}\approx\frac{g^{2}_{n}}{4}\sum^{\infty}_{m=-\infty}J^{2}_{m}\int^{\infty}_{-\infty}\frac{P_{\infty}(x_n)}{\gamma_n-i(\Delta_n-x_n+m\Omega)}dx_n,\nonumber\\
	&\braket{C_n(t_1)C^{\ast}_n(t_2)}\approx\frac{g^{4}_{n}}{16}\sum^{\infty}_{m,m'=-\infty}J^{2}_{m}J^{2}_{m'}\int^{\infty}_{-\infty}\frac{P_{\infty}(x_n)}{\left[\gamma_n+i(\Delta_n-x_n+m\Omega)\right]\left[\gamma_n-i(\Delta_n-x_n+m'\Omega)\right]}dx_n.
\end{align}
We next expand in the small quantity $x_n/\gamma_n$ to second order and keep only terms with $m=m'$. This gives
\begin{align}
	\label{eq:eq22}
	&\braket{C_n(t)}\approx\frac{g^{2}_{n}}{4}\sum^{\infty}_{m=-\infty}\frac{J^{2}_{m}}{\gamma_n-i(\Delta_n+m\Omega)}\left[1-\frac{\mu^{2}_{\rm{max},n}}{\left[\gamma_n-i(\Delta_n+m\Omega)\right]^2}\right],\nonumber\\
	&\braket{C_n(t_1)C^{\ast}_n(t_2)}\approx\frac{g^{4}_{n}}{16}\sum^{\infty}_{m=-\infty}\frac{J^{4}_{m}}{\gamma^{2}_n+(\Delta_n+m\Omega)^2}\left[1-\mu^2_{\rm{max},n}\frac{\gamma^{2}_n-3(\Delta_n+m\Omega)^2}{\left[\gamma^{2}_n+(\Delta_n+m\Omega)^2\right]^2}\right],
\end{align}
\end{widetext}
where we have used the fact that $\braket{x_n}=0$ (since $P_{\infty}(x_n)$ is a symmetric function) and $\braket{x^{2}_{n}}\approx\mu^{2}_{\rm{max},n}$. Hence to leading order in $\mu_{\rm{max},n}/\gamma_n$ we obtain
\begin{align}
	\label{eq:eq23}
	\braket{\delta C_n(t_1)\delta C^{\ast}_n(t_2)}\approx\frac{g^{4}_{n}}{16}\sum^{\infty}_{m=-\infty}\frac{J^{4}_{m}\mu^2_{\rm{max},n}}{\left[\gamma^{2}_n+(\Delta_n+m\Omega)^2\right]^2}.
\end{align}
Assuming $\delta<\gamma$ and taking into account the resonant contributions
we arrive at
\begin{align}
	-2 \ln D(t) \sim (\Gamma_{1,\rm{q}} t)^2\,\frac{\mu_{\rm max}^2}{\gamma^2} \,
	\frac{\delta}{\gamma}\,\sum_m J_m^4\ .
\end{align}
The effective dephasing rate, even in the absence of the modulation, is always smaller than the relaxation rate $\Gamma_{1,\rm{q}}$ and is further reduced by the modulations.

\bibliography{references}

\end{document}